\pdfoutput=1
\documentclass{JINST}
\usepackage{graphicx}
\usepackage{upgreek}
\title{The Resistive-WELL detector: a compact spark-protected single amplification-stage MPGD}

\author{G. Bencivenni$^a$\thanks{Corresponding author.}~, R. De Oliveira$^b$, G. Morello$^a$, M. Poli Lener$^a$\\
\llap{$^a$}Laboratori Nazionali di Frascati dell'INFN\\
Frascati, Italy\\
\llap{$^b$}CERN,
Meyrin, Switzerland\\
E-mail: \email{giovanni.bencivenni@lnf.infn.it}}
\abstract{In this work we present a novel idea for a compact spark-protected single amplification stage Micro-Pattern Gas Detector (MPGD).
The detector amplification stage, realized with a structure very similar to a GEM foil, is embedded through a resistive layer in the readout board. A cathode electrode, defining the gas conversion/drift gap, completes the detector mechanics.
The new structure, that we call Resistive-WELL (R-WELL), has some characteristics in common with previous MPGDs, such as C.A.T. and WELL, developed more than ten years ago. The prototype object of the present study has been realized  in the 2009 by TE-MPE-EM Workshop at CERN.
The new architecture is a very compact MPGD,  robust against discharges and exhibiting a large gain ($\sim$6$\times$10$^3$), simple to construct and easy for engineering and then suitable for large area tracking devices as well as huge calorimetric apparata.}


\begin{document}
\section{Introduction}
The modern photolitographic technology on flexible and standard PCB supports has allowed the invention of novel and robust MPGDs, such as GEM \cite{GEM1}, THGEM \cite{THGEM1},\cite{THGEM2} and Micromegas  \cite{MM1}.
These detectors exhibit good spatial \cite{ketzer} and time \cite{lhcb} resolution, high rate capability \cite{marco}, large sensitive area \cite{cms},
flexible geometry \cite{kloe2}, good operational stability \cite{lhcb2} and radiation hardness \cite{highrate}.\\
However, due to the fine structure and the typical micrometric distance of their electrodes, MPGDs generally suffer from spark occurrence that can eventually damage the detector.
The generation of the sparks in gas detectors is correlated with the transition from avalanche to streamer. This transition occurs when the Raether limit is overcome, that is when the primary avalanche size exceeds 10$^7$ - 10$^8$ ion-electron pairs \cite{Raether}.
In MPGDs, due to the very small distance between anode and cathode electrodes, the formation of the streamer can be easily followed by a discharge.\\
For GEMs the adopted solution is to share the gain among multiple amplification stages (double or triple-GEM structures), that allows a considerable reduction of the discharge rate \cite{discharge1},\cite{discharge2}.\\
For Micromegas the problem of the spark occurrence between the metallic mesh and the readout PCB has been solved with the introduction of a resistive layer deposition realized on top of the readout itself \cite{Peskov}. The principle is the same of the resistive electrode used in Resistive Plate Counters (RPCs) \cite{Pestov},\cite{santo},\cite{bencivenni}: the streamer, discharging a limited area around its location, is automatically quenched and the transition to spark is strongly suppressed giving the possibility to achieve large gains. \\
A further  limitation of such MPGDs is correlated with the complexity of their assembly procedure. In particular, a GEM chamber requires some time-consuming assembly steps such as the stretching and the gluing of the GEM foils \cite{Compass},\cite{LHCb},\cite{Totem}.  For this detector one of the co-authors of this paper has recently proposed  a solution based on the so called NS2 assembly approach  \cite{NS2}: an evolution of the stretching technique introduced for the GEM chambers of the LHCb experiment \cite{discharge2}.\\
Similar considerations can be also done for Micromegas, where the metallic mesh, defining the detector amplification region, requires a precise stretching.\\
%
The main goal of our project is the development of a novel MPGD  by combining in a unique approach the solutions and improvements realized in the last years in the MPGD field:
a very compact detector structure, robust against discharges and exhibiting large gains (up to 10$^4$), easy to build, cost effective and suitable for mass production. 
The novel detector, that we call Resistive-WELL (R-WELL), has some features  (such as electric field shape and signal formation) in common with some MPGDs developed by the end of last century (C.A.T. \cite{CAT} and WELL \cite{WELL}). The prototype discussed in this work, designed at the Laboratori Nazionali di Frascati and realized in the 2009 by TE-MPE-EM Workshop at CERN, has been developed in parallel with the CERN-GDD group \cite{BLIND-GEM},\cite{croci}.
\begin{figure}
  \begin{minipage}[t]{.46\textwidth}
    \centering
   \includegraphics[scale=0.28]{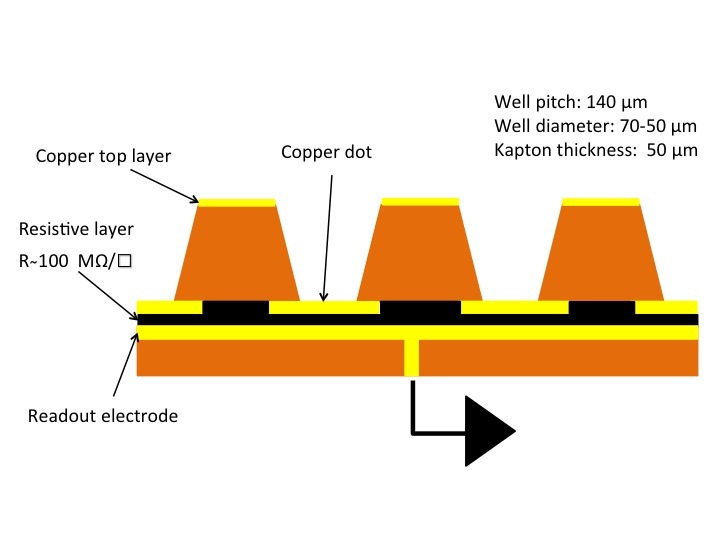}
    \caption{Schematic drawing of the R-WELL PCB.}
    \label{fig:blind-gem-substrate}
  \end{minipage}
    \hspace{5mm}
  \begin{minipage}[t]{.46\textwidth}
    \hspace{-7mm}
         \includegraphics[scale=0.3]{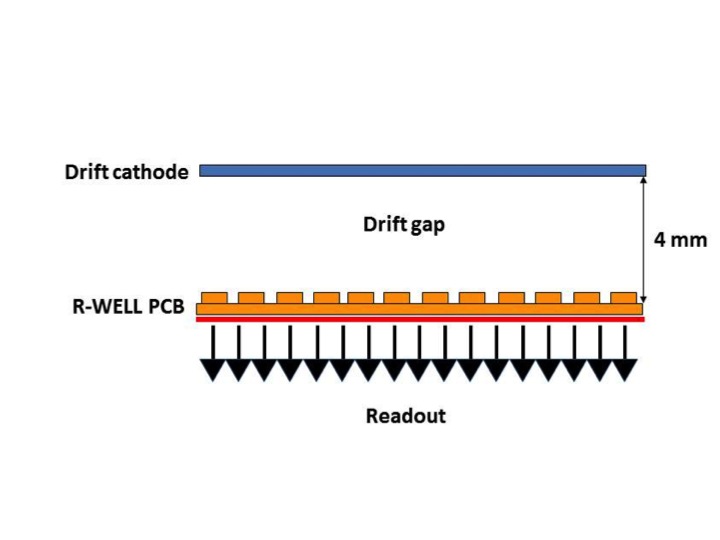}
         \caption{Schematic drawing of the R-WELL detector.}
    \label{fig:blind-gem-picture}
  \end{minipage}
\end{figure}
\section{Detector description}
\begin{figure}
  \begin{minipage}[t]{.46\textwidth}
    \centering
    \includegraphics[width=0.95\textwidth]{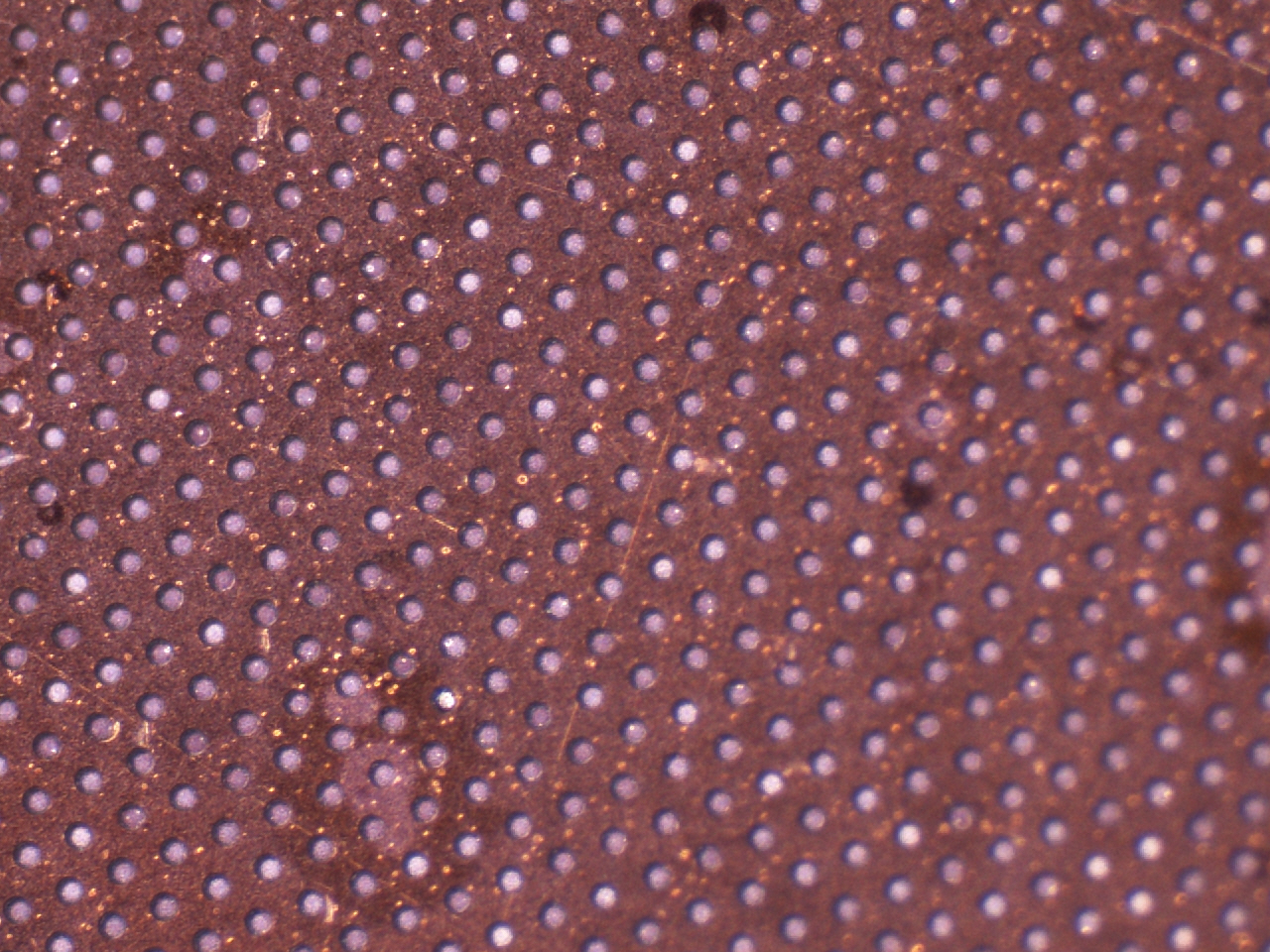}
    \caption{Microscope picture of the R-WELL PCB.}
    \label{fig:blind-GEM-proto}
  \end{minipage}
  \hfill
  \begin{minipage}[t]{.46\textwidth}
    \centering
    \includegraphics[width=0.99\textwidth]{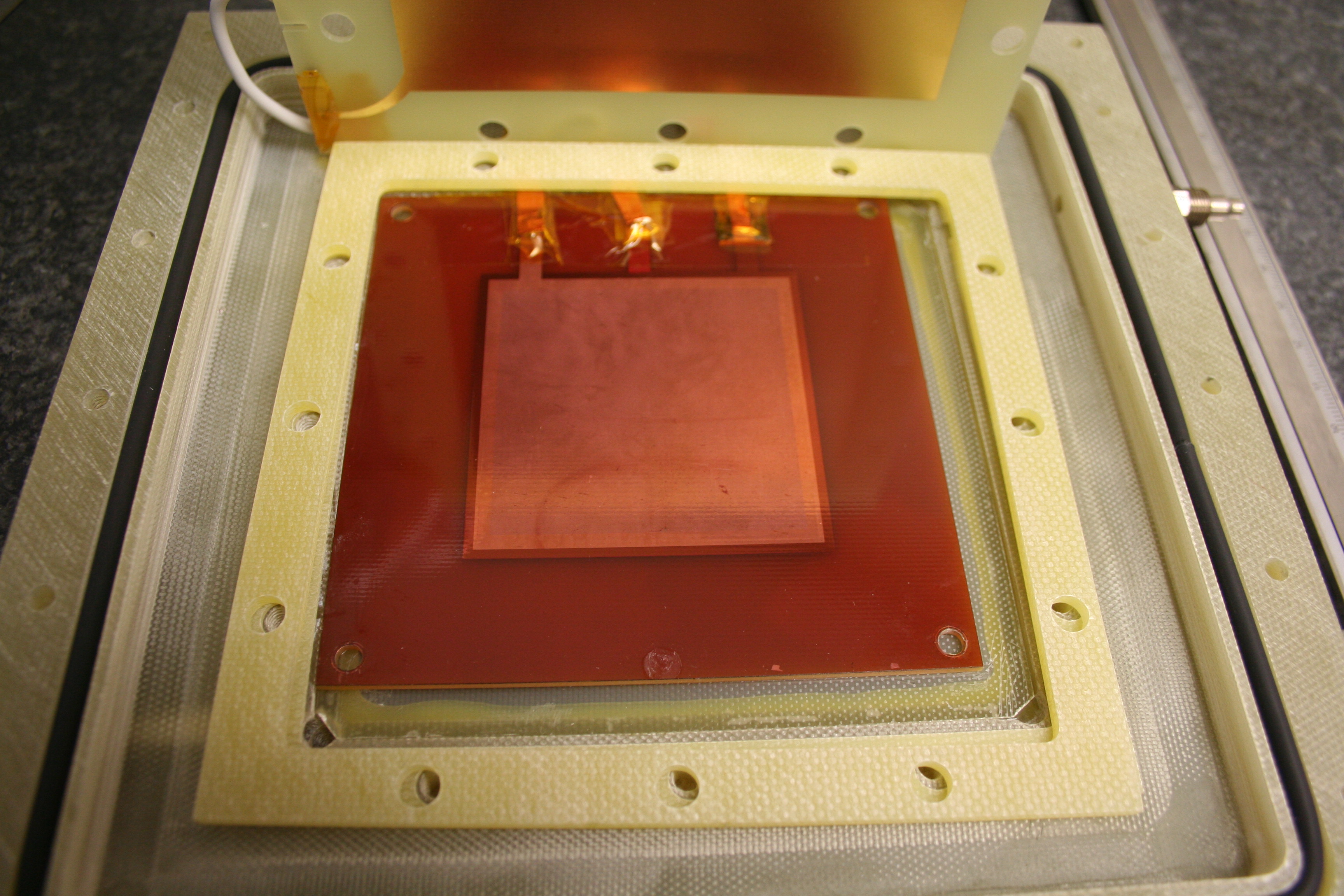}
    \caption{Detail of the R-WELL prototype.}
    \label{fig:blind-GEM-proto1}
  \end{minipage}
\end{figure}

The R-WELL prototype, as sketched in fig.~\ref{fig:blind-gem-substrate}, is realized by merging a suitable etched GEM foil with the readout PCB plane coated with a resistive deposition. The copper on the bottom side of the foil has been patterned in order to create small copper dots in correspondence of each WELL structure. The resistive coating has been performed by screen printing technique, more sophisticated sputtering technology such as Diamond Like Carbon (DLC) can be used for precise resistive layer patterning. 
The WELL matrix is hence realized on a 50$~\upmu$m thick polyimide foil, with conical channels 70$~\upmu$m (50$~\upmu$m) top (bottom) diameter and  140$~\upmu$m pitch. A cathode electrode, defining the gas conversion/drift gap, completes the detector mechanics (fig.~\ref{fig:blind-gem-picture}).\\
With respect to a classical single-GEM detector, the R-WELL is expected to exhibit a gas gain at least a factor of two larger.
Indeed in a  single-GEM detector only $\sim$50\% of the electron charge produced inside the holes contributes to the formation of the signal, while the rest of the electron charge is collected by the lower side of the GEM foil. In addition the signal in a GEM detector is mainly due to the electron motion, because the ion component is largely shielded by the GEM foil itself and the avalanche is  confined in the holes.
\begin{figure}
  \begin{minipage}[t]{.46\textwidth}
    \centering
    \includegraphics[width=1.\textwidth]{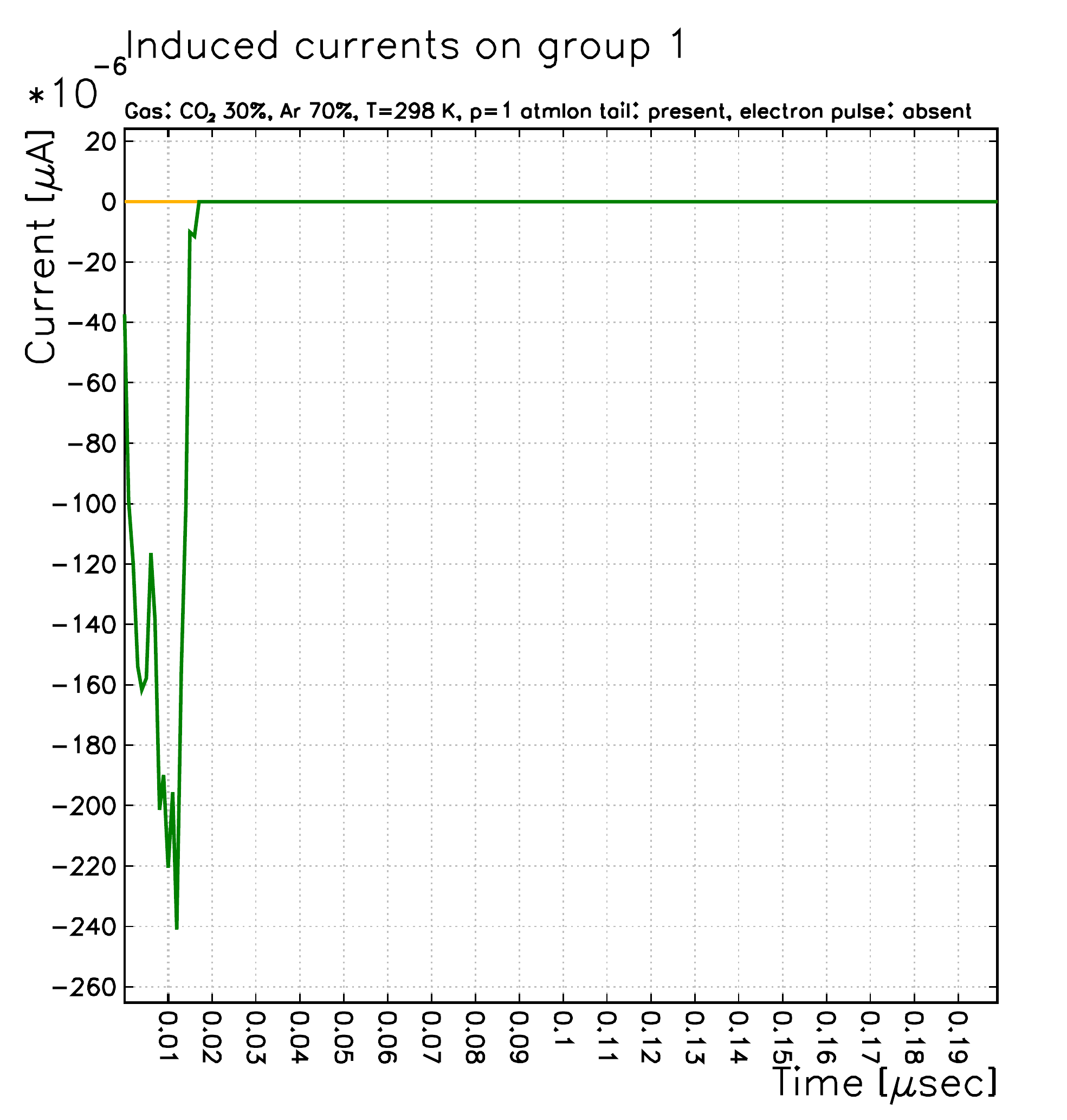}
    \caption{Simulation of a signal from a single ionization electron in a single-GEM detector in Ar/CO$_2$ = 70/30 gas mixture. The duration of the signal, about 20 ns, depends on the induction gap thickness, drift velocity and electric field in the gap. }
    \label{fig:GEM-signal}
  \end{minipage}
  \hfill
  \begin{minipage}[t]{.46\textwidth}
    \centering
    \includegraphics[width=1.\textwidth]{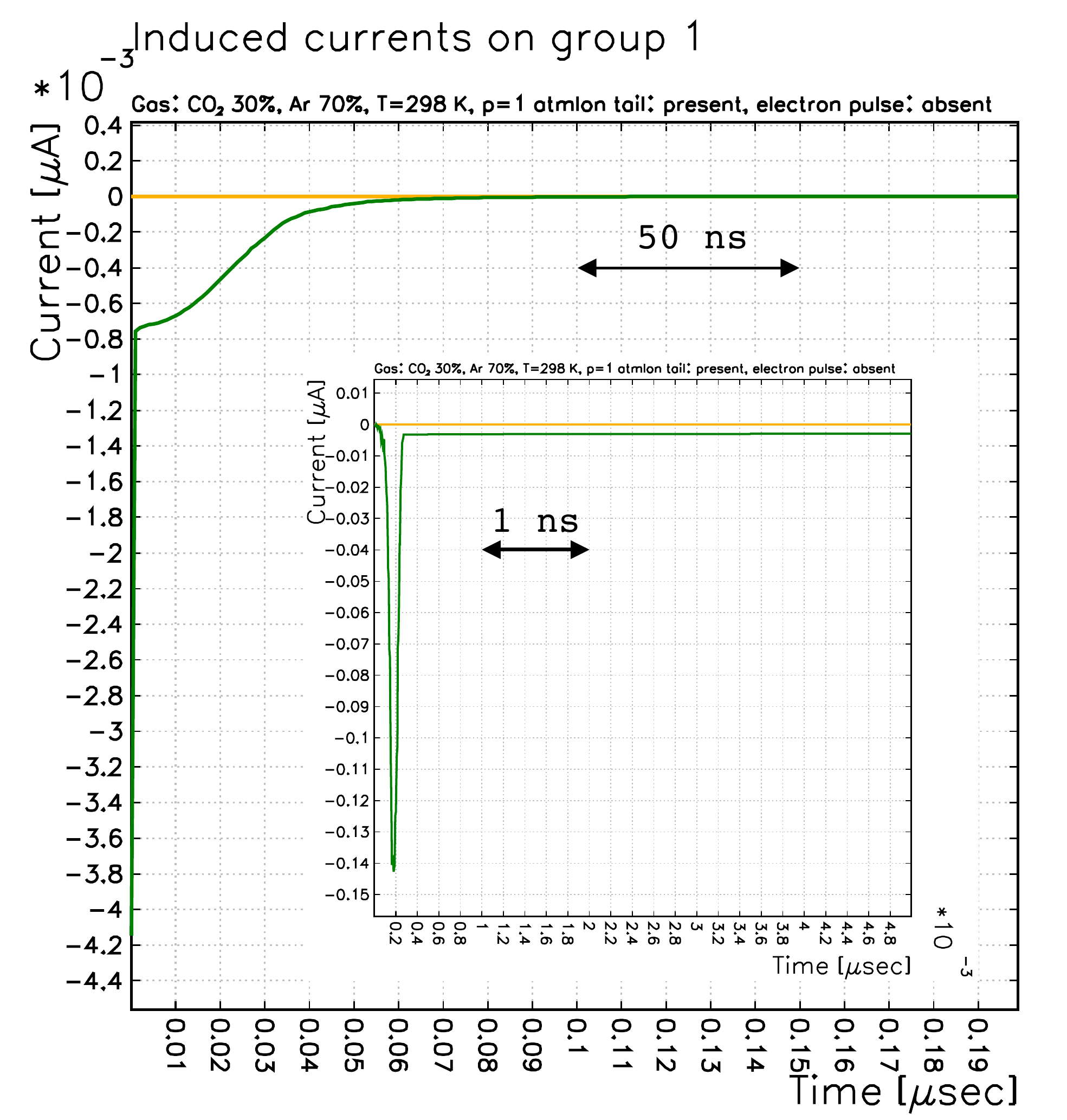}
    \caption{Simulation of a signal from a single ionization electron in a R-WELL in Ar/CO$_2$ = 70/30 gas mixture. The absence of the induction gap is responsible for the fast initial spike, about 200 ps, induced by the motion and fast collection of the electrons and followed by a 50 ns ion tail. }
    \label{fig:B-GEM-signal}
  \end{minipage}
\end{figure}\\
In a R-WELL the whole electron charge produced into the amplification channel is promptly collected on the resistive layer (capacitively coupled with the readout plane) through the copper dot. Moreover also the ionic component, apart ballistic effects correlated with the integration time of the readout electronics, contributes to the formation of the signal in a similar way as the electron part.
In fig.~\ref{fig:GEM-signal} and fig.~\ref{fig:B-GEM-signal} a comparison between the simulation of the signal generated by a single ionization electron in a single-GEM and a R-WELL is reported. In the R-WELL a further increase of the gain could be achieved thanks to the resistive electrode which, quenching the discharges, allows a higher amplification field inside the channel.\\
%
A distinctive advantage of the proposed technology is that the detector, composed by very few components, does not require complex and time-consuming assembly procedures: neither stretching, nor gluing, nor internal support frames are required. 
%
The prototype described in this paper has a 5$\times$5 cm$^2$ active area and the layer surface resistivity is $\sim$100 $\rm{M}\Omega/\square$. The coating of the PCB has been realized by a polymer resistive paste deposition using the screen printing technique. The readout electrode has been realized as a unique pad as large as the detector active area. In figg.~\ref{fig:blind-GEM-proto},~\ref{fig:blind-GEM-proto1} some details of the R-WELL PCB and the prototype are respectively shown.
\section{Detector performance}
\begin{figure}
\vspace{-.5cm}
\centering
 \begin{minipage}[t]{.65\textwidth}
    \includegraphics[width=1.1\textwidth]{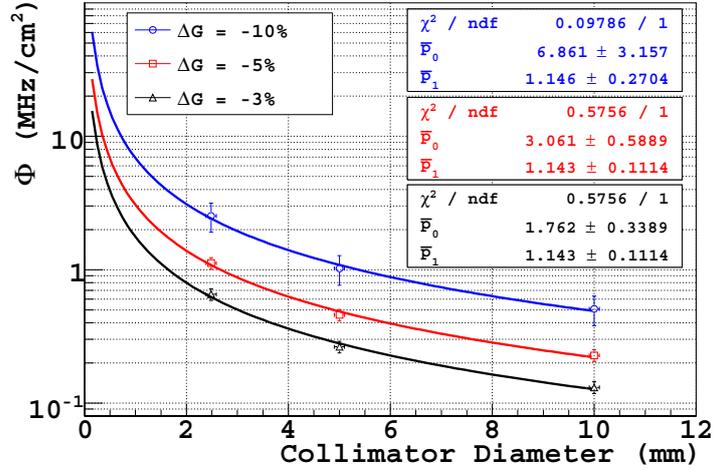}
    \vspace{-1.cm}
    \caption{Rate capability (with X-rays) for the R-WELL as a function of the diameter of the collimator, for different value of the accepted gain drop (-3 \% black line; -5\% red; -10\% blue).}
   \label{fig:rate-capability-3}
\end{minipage}	
\end{figure}

The detector has been characterized  by measuring the gas gain, rate capability and discharge behaviour in current mode, that is by recording the current drawn by the resistive layer under irradiation.
The device has been irradiated with a collimated flux of 5.9 keV X-rays generated by a  PW2217/20 Philips.
The detector gain has been measured as a function of the potential applied between the top electrode of the amplification stage and the resistive layer (see fig.~\ref{fig:blind-gem-substrate}). For the measurement a Keithley-6485 pico-ammeter with 10 fA sensitivity  has been used.
\begin{figure}
  \begin{minipage}[t]{.48\textwidth}
   \centering
    \includegraphics[width=1.05\textwidth]{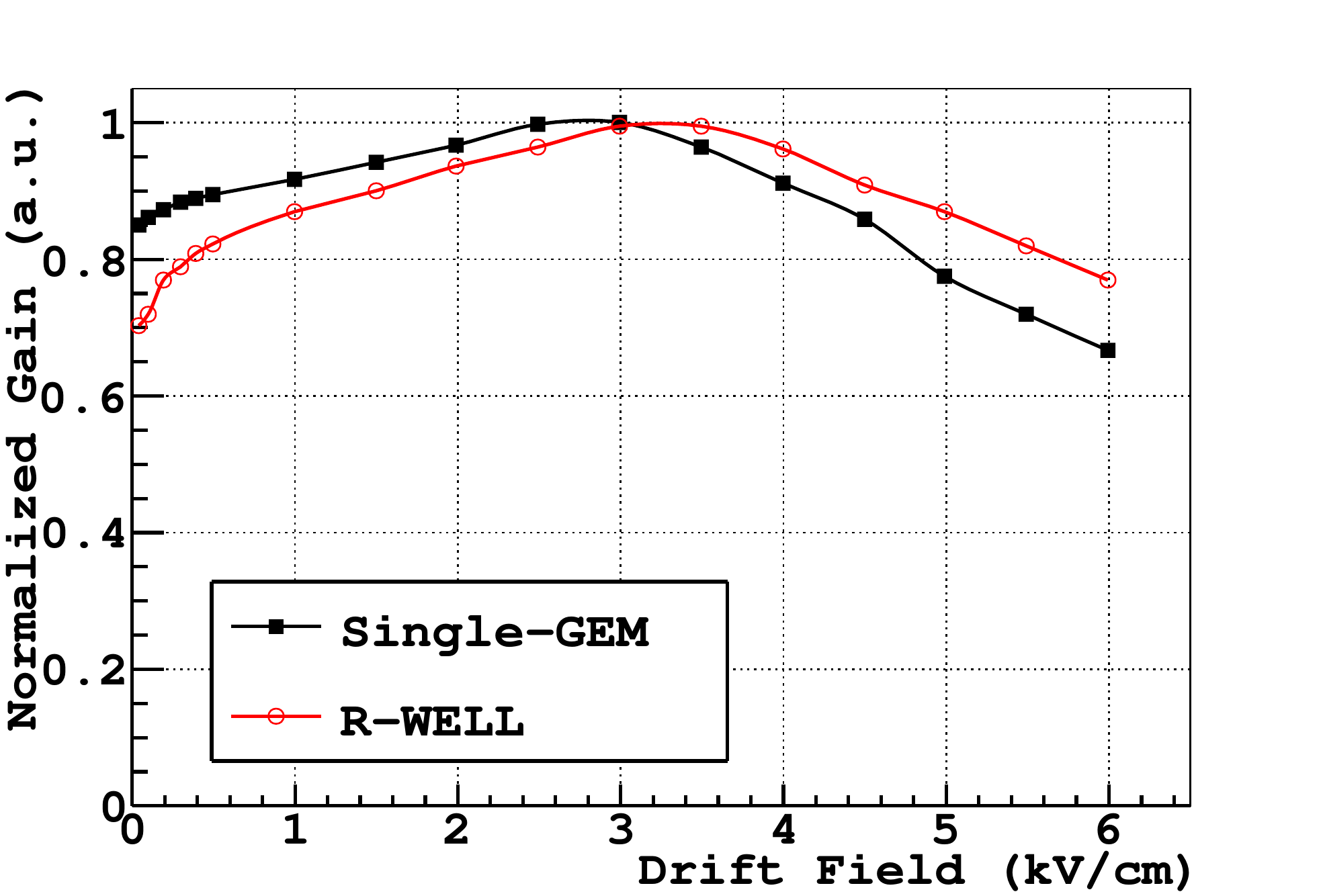}
    \caption{Relative charge collection efficiency as a function of the drift field.}
    \label{fig:drift_field}
  \end{minipage}
  \hfill
  \begin{minipage}[t]{.48\textwidth}
    \centering
   \includegraphics[width=1.05\textwidth]{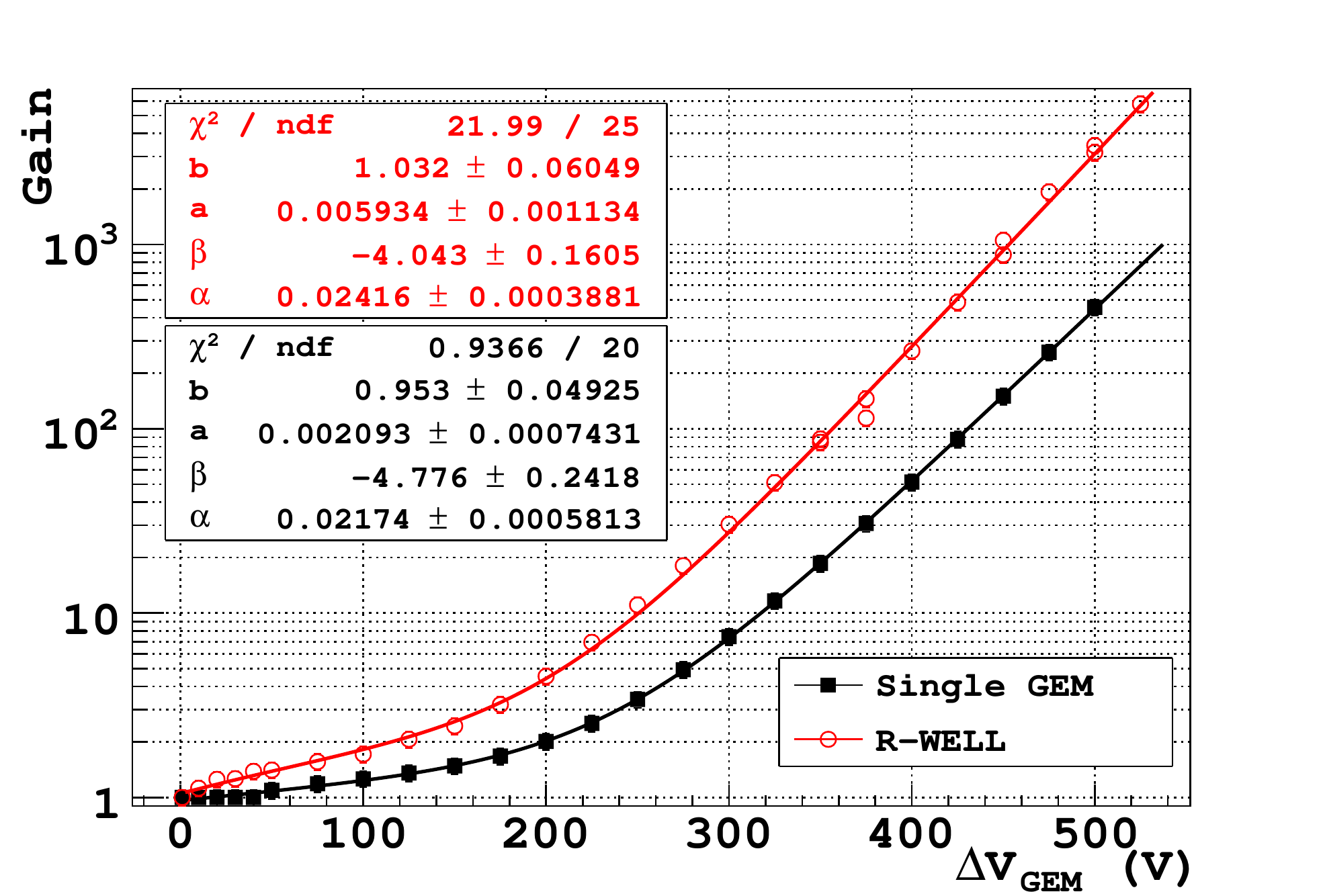}
   \caption{Gas gain for the R-WELL (red points) and the single-GEM (black points).}
   \label{fig:gain-comparison-1}
  \end{minipage}
\end{figure}\\
In a structure like the one discussed in this paper (as well as for a GEM) some of the field lines in the conversion/drift region are expected to terminate on the metallized layer of the upper part of the amplification stage. Primary electrons following these lines are not collected into the WELL (hole) and therefore are not multiplied, producing a collection inefficiency. As shown in fig.~\ref{fig:drift_field} the collection efficiency (estimated from the normalized gain \cite{bible}), for a fixed value of the voltage applied to the WELL (hole) structure, depends on the drift field. For the prototype studied in this work the maximum collection efficiency has been found in correspondence of a drift field of about 3.5 kV/cm ($<$3.0 kV/cm for the single-GEM).
As reported in fig.~\ref{fig:gain-comparison-1} the maximum gain achievable with the R-WELL (G$\sim$6000 at $\Delta$V$=$525 V) is significantly higher than that one exhibited by a standard single-GEM used as reference (G$<$1000  at $\Delta$V$=$500 V)\footnote{The gain has been parametrized as follows:
$G(V) = b+a\cdot V~~\rm{for~V<225},~~~
G(V) =  e^{\beta+\alpha\cdot V~~}\rm{for~V>225}$}.
%
%
\begin{figure}
  \begin{minipage}[t]{.48\textwidth}
   \centering
    \includegraphics[width=1.05\textwidth]{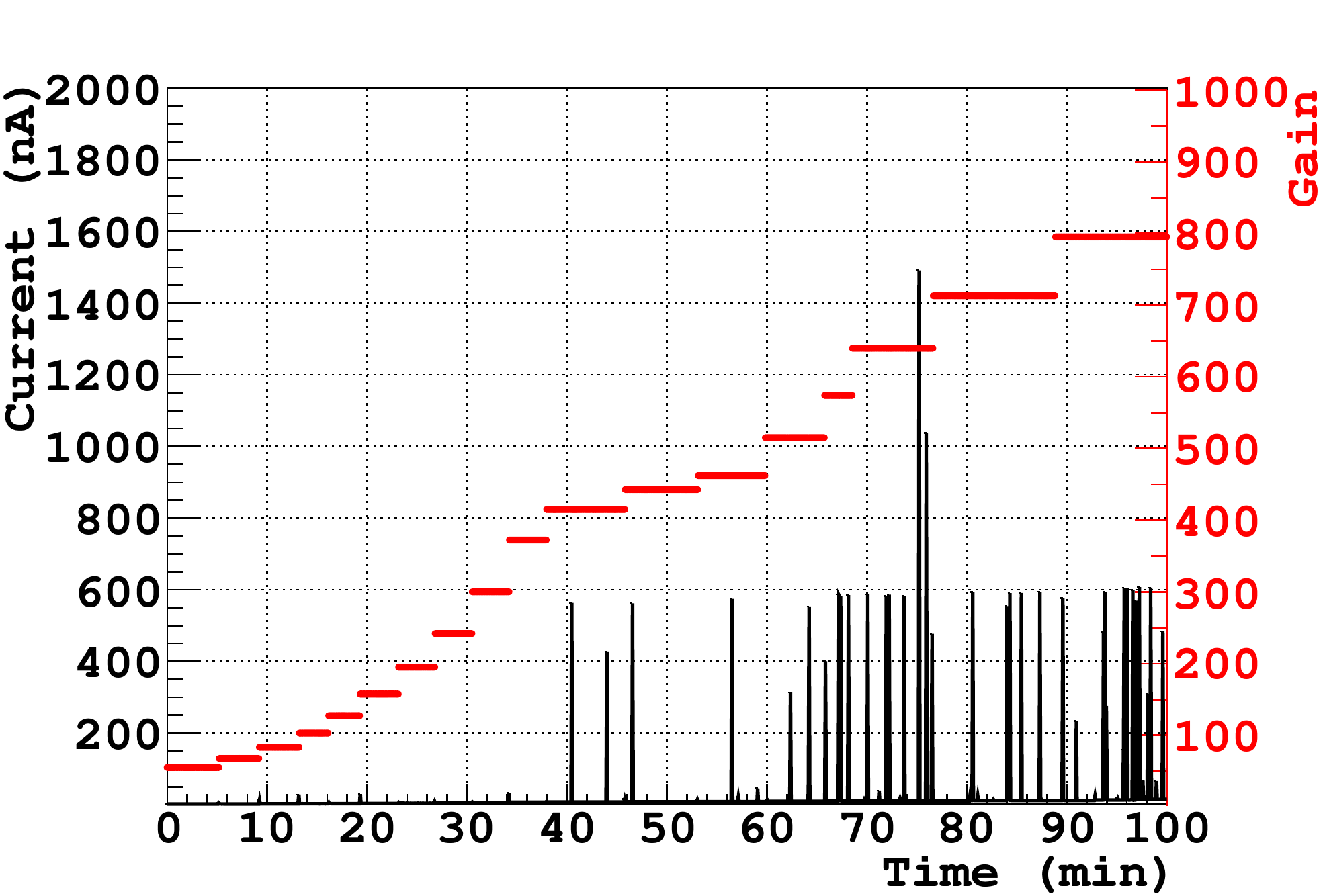}
    \caption{Monitoring of the current drawn by the single-GEM detector for different gas gain. Discharge amplitudes as high as 1$\upmu$A are recorded at higher gains.}
    \label{fig:discharge_gem}
  \end{minipage}
  \hfill
  \begin{minipage}[t]{.48\textwidth}
    \centering
   \includegraphics[width=1.05\textwidth]{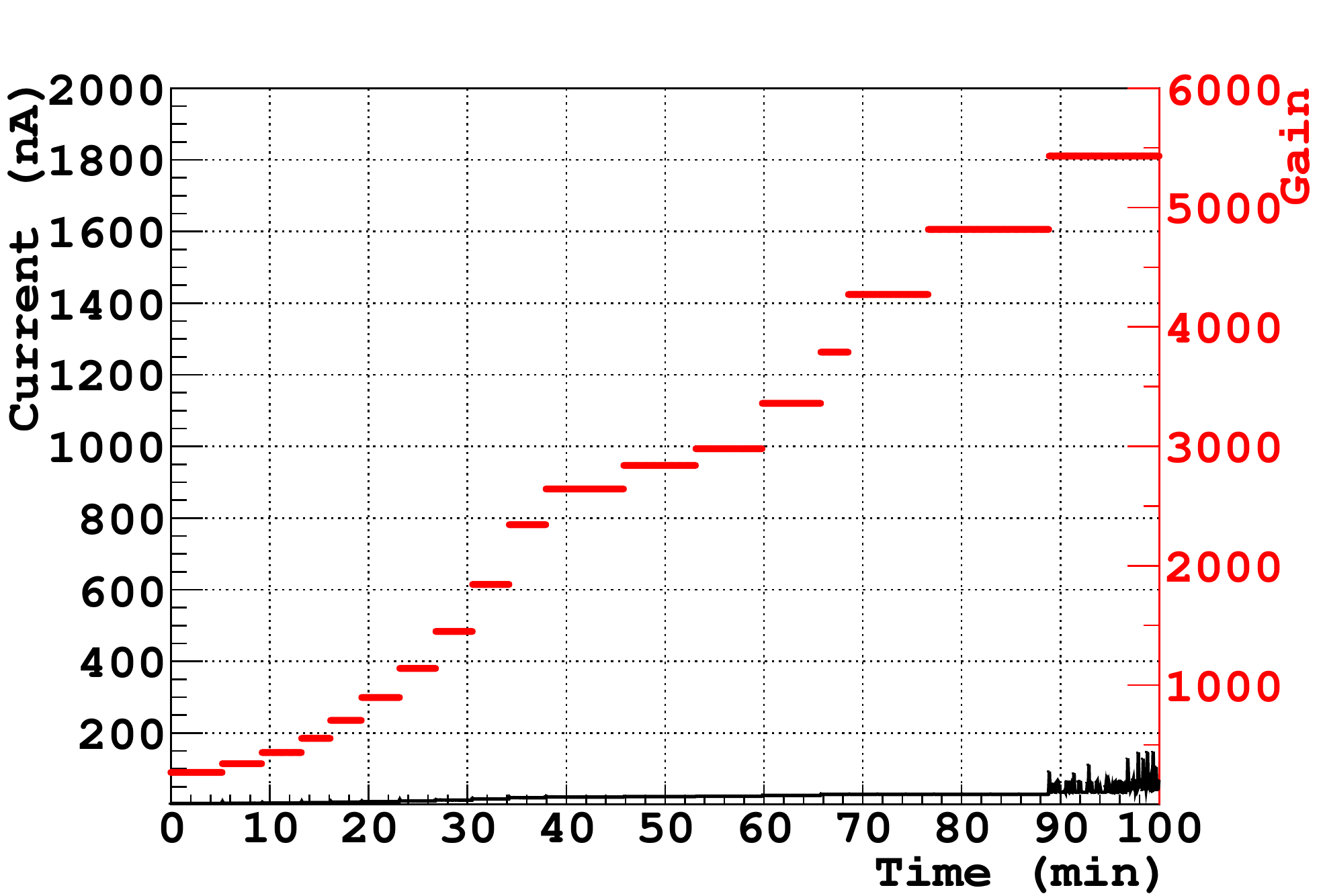}
  \caption{Monitoring of the current drawn by the R-WELL detector for different gas gain. Discharges are quenched down to few tens of nA even at high gains.}
   \label{fig:discharge_R-WELL}
  \end{minipage}
\end{figure}\\
For both detectors the maximum voltage achieved during the measurement is correlated with the onset of the discharge activity, that, from a very qualitative analysis, comes out to be substantially different for the two devices.
As shown in fig.~\ref{fig:discharge_gem} and fig.~\ref{fig:discharge_R-WELL}, the typical  discharge amplitude for the R-WELL is of the order of few tens of nA (in anycase less than 100 nA also at the maximum gain), while for the GEM discharges with amplitude of the order of $\upmu$A are observed at high gas gain. Further systematic and more quantitative studies on this item must be clearly performed.
\begin{figure}
  \begin{minipage}[t]{.48\textwidth}
    \centering
    \includegraphics[width=1.1\textwidth]{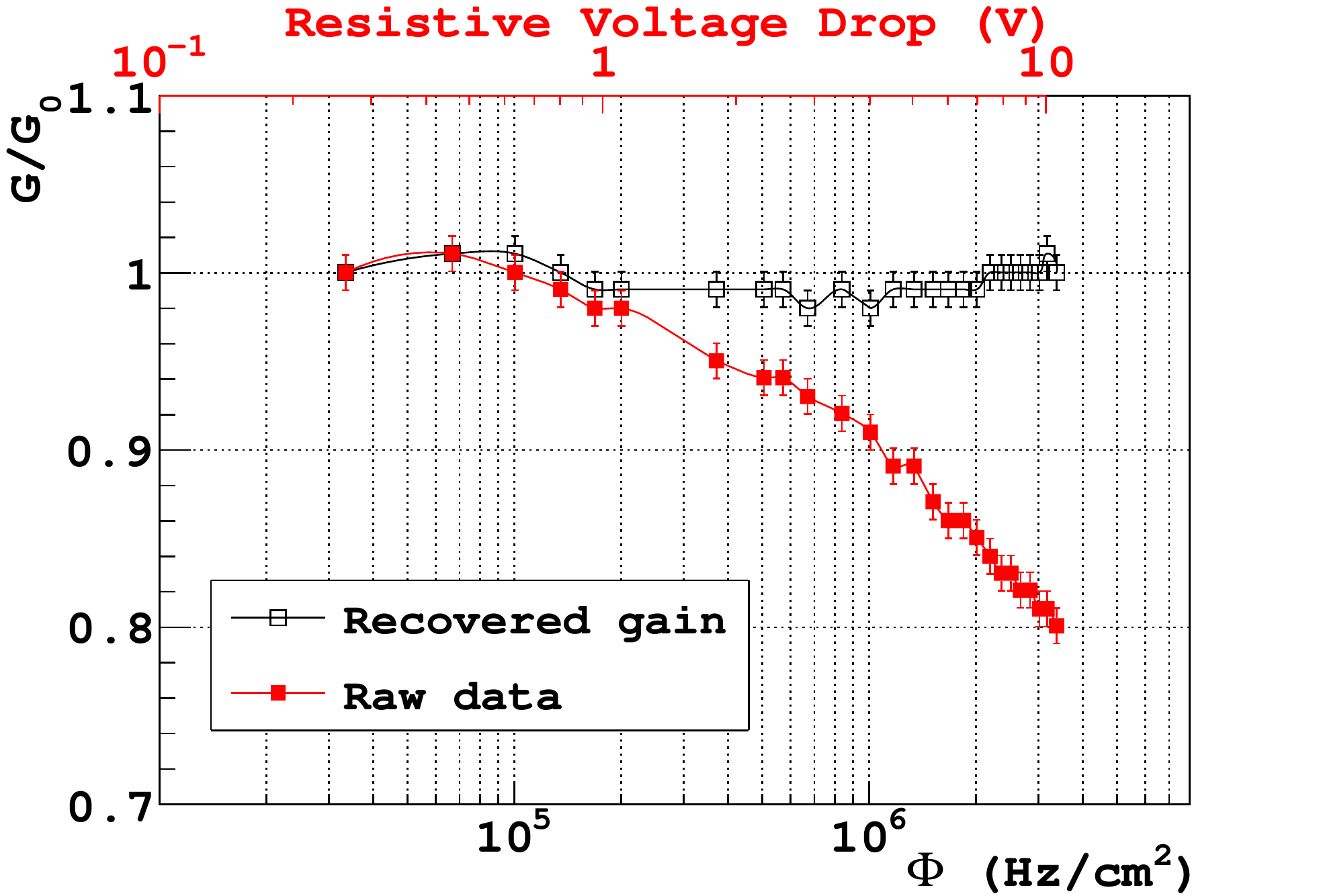}
     \caption{Normalized gain (a.u.) for the R-WELL as a function of the flux: full squares are the raw data; open squares are obtained increasing the voltage (of a value reported on the upper horizontal axis) in order to recover the gain.}
     \label{fig:rate-capability-2}
  \end{minipage}
  \hfill
  \begin{minipage}[t]{.48\textwidth}
     \centering
     \includegraphics[width=1.1\textwidth]{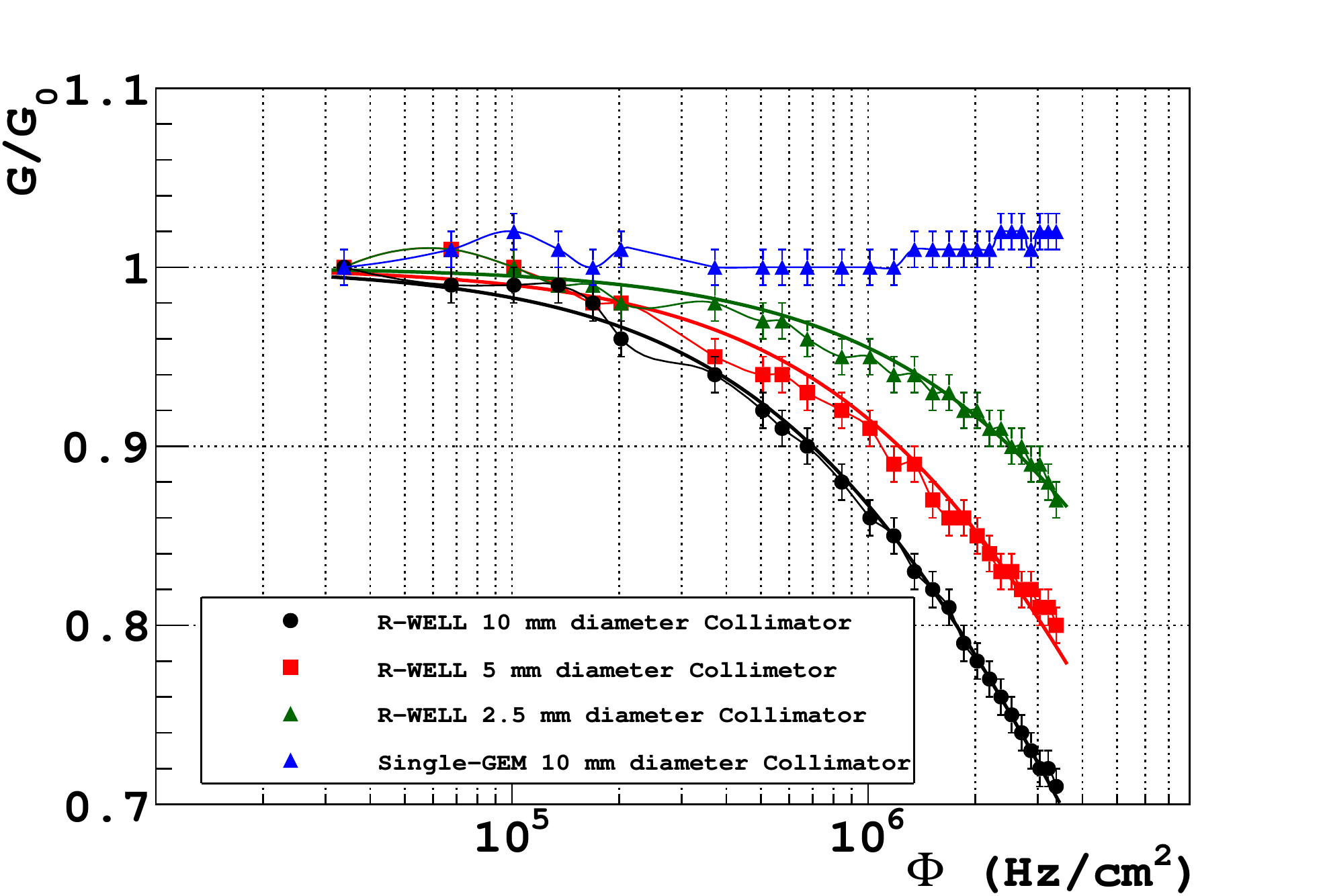}
    \caption{Comparison of the normalized gain (a.u.) for the GEM (blue) and the R-WELL for different collimator diameters (10 mm - black; 5 mm - red; 2.5 mm - green).}
    \label{fig:rate-capability-1}
  \end{minipage}
\end{figure}\\
On the other hand a drawback correlated with the  introduction of a high resistivity layer between the amplification stage and the readout is the reduced capability to stand high particle fluxes: larger the radiation rate, higher is the current drawn through the resistive layer and, as a consequence, larger the drop of the amplifying voltage (fig. \ref{fig:rate-capability-2}). 
%
%
In fig.~\ref{fig:rate-capability-1} the normalized gain of the single-GEM is compared with that obtained for the R-WELL with three different collimator diameters: 10 mm, 5 mm and 2.5 mm. The gain of the GEM is substantially constant over the explored  range of  radiation flux (up to 3 MHz/cm$^2$), while the maximum particle flux that the R-WELL is able to stand, in agreement with an Ohmic behavior of the detector (see appendix \ref{app:drop}), decreases with the increase of the effective diameter of the X-ray spot on the detector. The points in fig. \ref{fig:rate-capability-1} are fitted with the function
\begin{equation}
\frac{G}{G_{0}}=\frac{-1+\sqrt{1+4p_{0}\Phi}}{2p_{0}\Phi}
\label{fit}
\end{equation}
for reasons explained in appendix \ref{app:drop}. 
%
%
The function \ref{fit} has been used to evaluate the radiation flux when the detector is expected to have a gain drop of 3\%, 5\% and 10\% for all the collimators. These results, as shown in fig. \ref{fig:rate-capability-3}, seem to indicate that the rate capability of the detector, for a fixed surface resistivity, can be tuned with a suitable segmentation of the resistive layer by means a conductive grid or with an equivalent current evacuation scheme.
In addition, taking into account the higher ionization of the X-rays used in this measurement ($\sim$210 electron-ion pairs), for m.i.p. ($\sim$30 electron-ion pairs in 4 mm drift gap) the rate capability is expected to be larger of a factor $\sim$7. As a consequence, with a proper segmentation of the resistive layer, a rate capability of $\sim$ 1 MHz/cm$^2$ for m.i.p seems to be achievable. 
%
\section{Conclusions}
The R-WELL detector investigated in this work shows several advantages: small thickness (few millimeters), effective spark quenching, very simple assembly procedure, besides of good gas gain  (G$\sim$6000) and rate capability ranging for X-rays from 100 kHz/cm$^2$ to 600 kHz/cm$^2$ (G$\sim$2000 and a surface re\-si\-sti\-vi\-ty of about  100 M$\Omega/\square$). Larger gas gain could be achieved using thicker kapton foils for the realization of the R-WELL structure (polyimide foils with a thickness up to 125 $\upmu$m are produced from KANEKA ltd in Japan).  Higher rate capability could be obtained using a suitable segmentation of the resistive layer and adjusting its surface resistivity. All these features make the R-WELL detector a valuable solution for large size tracking and calorimetric apparata where reliability, construction simplicity and cost-effective technology are recommendable.
\appendix
\section{The gain drop in a R-WELL}
\label{app:drop}
The experimental data reveal that the gain variation of a R-WELL depends on the radiation flux (fig. \ref{fig:rate-capability-1}) and the observed drop is supposed to be due to the resistive layer of the detector as explained in the following. The gain of a R-WELL can be written as a function of the potential difference applied on the amplification stage:
\begin{equation}
G_{0}=e^{\beta+\alpha\cdot V_{0}}
\end{equation}
A gain drop corresponds to a decrease of the voltage $V_{0}$
\begin{eqnarray}
G & = & e^{\beta+\alpha\cdot\left(V_{0}-\delta V\right)}\nonumber \\
{} & = & G_{0}e^{-\alpha\cdot\delta V}
\label{drop} 
\end{eqnarray}
Assuming that the gain drop is only due to the resistive layer we can write the Ohm's first law
\begin{equation}
\delta V = i \cdot\Omega
\end{equation}
where $i$ is the current measured on the resistive layer and $\Omega$ is the \emph{average} resistance faced by the charges to reach the ground frame. The current $i$ can be written as follows
\begin{eqnarray}
i & = & eN_{0}GR\nonumber 
\end{eqnarray} 
with $e$ the electron charge, $N_{0}$ is the number of electrons produced by a single $5.9~\rm{keV}$ photon (from GARFIELD simulations we have $N_{0}=209$), $G$ is the gain of the GEM foil and $R$ is the conversion rate of the X-ray in the gas. The rate can be also expressed in terms of a flux $\Phi$ standing the simple relation $R=\Phi\pi r^{2}$, with $r$ the collimator radius.
Eq. \ref{drop} becomes
\begin{eqnarray}
G & = & G_{0}e^{-\alpha e N_{0}G\Phi\pi r^{2}\Omega}\nonumber\\
\frac{G}{G_{0}}e^{\alpha e N_{0}G\Phi\pi r^{2}\Omega} & = & 1\nonumber
\end{eqnarray}
The exponential can be expanded using the Maclaurin series up to the first order since the exponent is smaller than 1.
\begin{eqnarray}
\frac{G}{G_{0}}\left[1+\alpha e N_{0}G\Phi\pi r^{2}\Omega\right]&=&1 \nonumber
\end{eqnarray}
re-writable as
\begin{eqnarray}
\alpha e N_{0}G_{0}\Phi\pi r^{2}\Omega\left(\frac{G}{G_{0}}\right)^2+\frac{G}{G_{0}}-1&=&0\nonumber
\label{formula}
\end{eqnarray}
We insert the parameter $p_{0}$ defined as 
\begin{equation}
p_{0}=\alpha e N_{0}G_{0}\Omega\pi r^{2}
\label{p0}
\end{equation}
which lets us write the solution of eq. \ref{formula} in a very simple way:
\begin{equation}
\frac{G}{G_{0}}=\frac{-1+\sqrt{1+4p_{0}\Phi}}{2p_{0}\Phi}
\label{solution}
\end{equation}
where the only positive solution is taken since the argument under square root is larger than 1. \\
Eq. \ref{solution} gives the ratio of $G$ to $G_{0}$ as a function of the X-rays flux and it has been used to fit the graphs in fig. \ref{fig:rate-capability-1}.

\section{A model for the resistance $\Omega$: circular approximation}
\begin{figure}[!h]
  \begin{center}
	\includegraphics[scale=0.55]{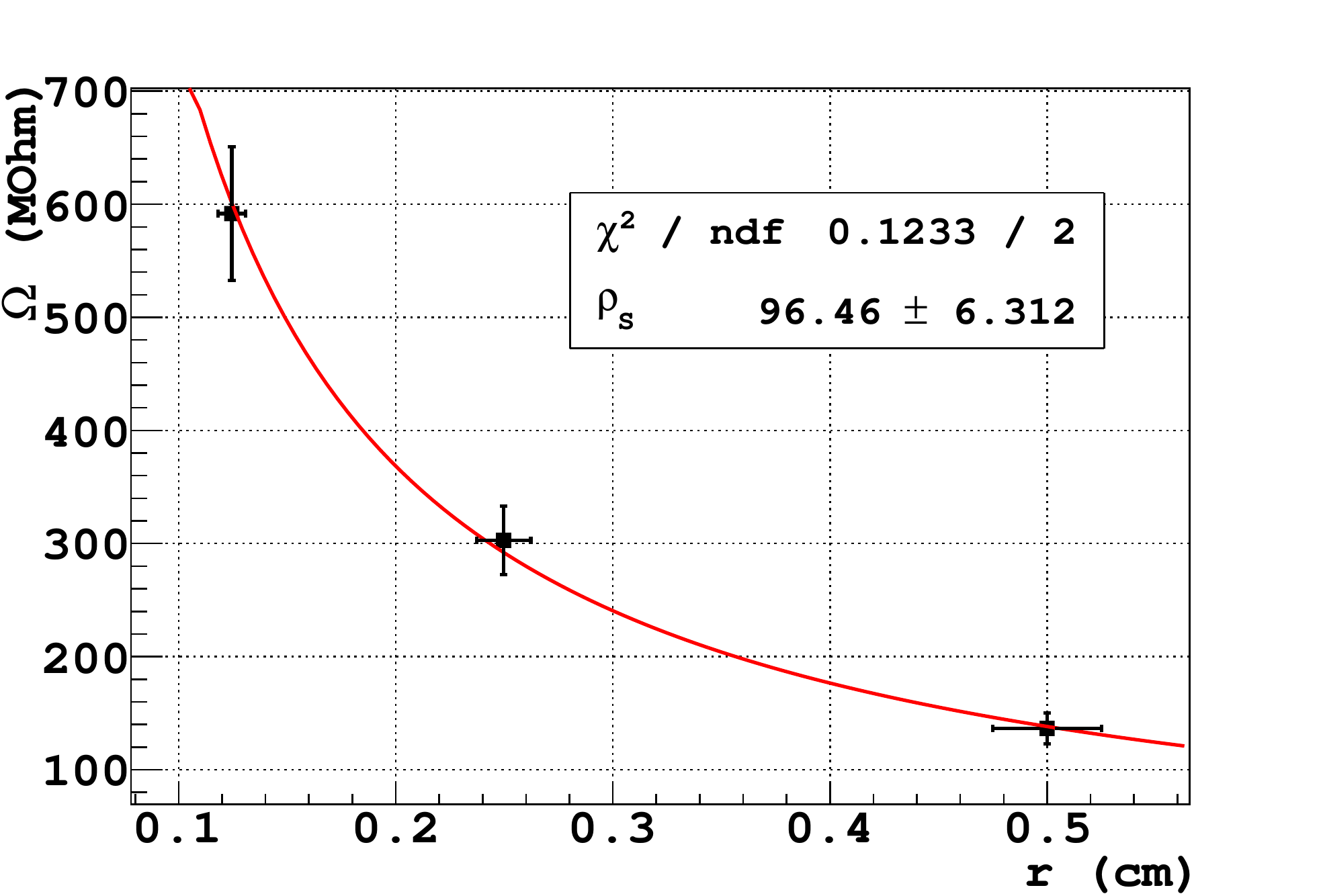}
	\caption{The \emph{average} resistance $\Omega$ as a function of the collimator radius $r$.}
  \label{plot_res}
\end{center}
\end{figure}

According to eq. \ref{p0}, the relation between the parameter $p_{0}$ and the \emph{average} resistance $\Omega$ is given by 
\begin{eqnarray}
\Omega(r) &=& \frac{p_{0}(r)}{\alpha e N_{0}G_{0}\pi r^{2}}\nonumber
\end{eqnarray}
The dependence on the collimator radius $r$ is shown in fig. \ref{plot_res}. We propose in the following a simplified model exploiting this dependence.\newline
Let's consider the axis of the collimator pointing to the center of the resistive layer and let $d$ be the half of side length of the active area and $r$ the collimator radius. We consider the case $r\ll d$, so that we can approximate the active area as a circle with radius $d$. The charge produced at a distance $\xi \in\left[0,r\right]$ from the center will cover the path $d-\xi$. The average path can be evaluated as follows: 
\begin{eqnarray}
<d-\xi> &= &\frac{\int_{0}^{r}\int_{0}^{2\pi}\left(d-\xi\right) \rm{d}\xi \rm{d}\theta}{\int_{0}^{r}\int_{0}^{2\pi}\rm{d}\xi \rm{d}\theta} \nonumber \\
{} &=&d-\frac{r}{2}
\label{b1}
\end{eqnarray} 
So we can consider that all the charge produced in the area $\pi r^2$ is concentrated in the circle with radius $r/2$. From that point the charge moves towards the ground frame ideally in a pipe with height $\delta$ (the thickness of the resistive layer) and width $r\rm{d}\theta$, with this width constant all over the path. Under this assumption the total surface $S$ crossed by all the charge is
\begin{equation}
S = \delta\cdot\int_{0}^{2\pi}\frac{r}{2}\rm{d}\theta=\delta\cdot\pi r
\label{b2}
\end{equation}
According to the second Ohm's law, using the results in \ref{b1} and \ref{b2}, the \emph{average} resistance $\Omega$ is
\begin{eqnarray}
\Omega &=& \rho_{V}\frac{d-\frac{r}{2}}{\delta\cdot\pi r}\nonumber \\
&=& \rho_{S}\frac{d-\frac{r}{2}}{\pi r}
\label{resistenza}
\end{eqnarray}
where we used the definition of surface resistivity $\rho_{S}=\rho_{V}/\delta$.
The points in fig. \ref{plot_res} are fitted with the function \ref{resistenza} having used $\rho_{S}$ as parameter of the fit. In this case we obtain $\rho_{S}=96\pm6~\rm{M}\Omega/\square$ compatible with the value of $100~\rm{M}\Omega/\square$ declared by the deliverer (and co-author of this paper) of the detector PCB.

\end{document}